\documentclass[]{aastex631}
\usepackage{graphicx}%
\usepackage{multirow}%
\usepackage{amsmath,amssymb,amsfonts}%
\usepackage{amsthm}%
\usepackage{mathrsfs}%
\usepackage{xcolor}%
\usepackage{rotating}

\newcommand{\nustar}{\textit{NuSTAR}}
\newcommand{\chandra}{\textit{Chandra}}
\newcommand{\fermi}{\textit{Fermi}-LAT}

\begin{document}

\title{Spectrum and location of ongoing extreme particle acceleration in Cassiopeia A}

\correspondingauthor{Jooyun Woo}
\email{jw3855@columbia.edu}

\author[0009-0001-6471-1405]{Jooyun Woo}
\affiliation{Columbia Astrophysics Laboratory, Columbia University, 538 West 120th Street, New York, NY 10027, USA}

\author[0000-0002-9709-5389]{Kaya Mori}
\affiliation{Columbia Astrophysics Laboratory, Columbia University, 538 West 120th Street, New York, NY 10027, USA}

\author[0000-0002-3681-145X]{Charles J. Hailey}
\affiliation{Columbia Astrophysics Laboratory, Columbia University, 538 West 120th Street, New York, NY 10027, USA}

\author[0009-0009-6387-7241]{Elizabeth Spira-Savett}
\affiliation{Department of Physics and Astronomy, Barnard College, 3009 Broadway, New York, NY 10027, USA}

\author[0000-0003-0890-4920]{Aya Bamba}
\affiliation{Department of Physics, Graduate School of Science, The University of
Tokyo, 7-3-1 Hongo, Bunkyo-ku, Tokyo, 113-0033, Japan}
\affiliation{Research Center for the Early Universe, School of Science, The
University of Tokyo, 7-3-1 Hongo, Bunkyo-ku, Tokyo, 113-0033, Japan}
\affiliation{Trans-Scale Quantum Science Institute, The University of Tokyo, 7-3-1
Hongo, Bunkyo-ku, Tokyo, 113-0033, Japan}

\author[0000-0002-1984-2932]{Brian W. Grefenstette}
\affiliation{Space Radiation Lab, California Institute of Technology, 1200 East
California Boulevard, Pasadena, CA 91125, USA}

\author[0000-0002-1432-7771]{Thomas B. Humensky}
\affiliation{Goddard Space Flight Center, NASA, 8800 Greenbelt Road, Greenbelt,
MD 20771, USA}
\affiliation{Department of Physics, University of Maryland, 4150 Campus Drive,
College Park, MD 20742, USA}

\author[0000-0002-3223-0754]{Reshmi Mukherjee}
\affiliation{Department of Physics and Astronomy, Barnard College, 3009 Broadway, New York, NY 10027, USA}

\author[0000-0001-6189-7665]{Samar Safi-Harb}
\affiliation{Department of Physics and Astronomy, University of Manitoba, 30A
Sifton Road, Winnipeg, MB R3T 2N2, Canada}

\author[0000-0001-7380-3144]{Tea Temim}
\affiliation{Department of Astrophysical Sciences, Princeton University, 4 Ivy
Lane, Princeton, NJ 08544, USA}

\author[0000-0001-7209-9204]{Naomi Tsuji}
\affiliation{Faculty of Science, Kanagawa University, 3-27-1 Rokukakubashi,
Kanagawa-ku, Yokohama-shi, Kanagawa, 221-8686, Japan}



\begin{abstract}

Young supernova remnants (SNRs) are believed to be the origin of energetic cosmic rays (CRs) below the ``knee" of their spectrum at $\sim3$ petaelectronvolt (PeV, $10^{15}$ eV). Nevertheless, the precise location, duration, and operation of CR acceleration in young SNRs are open questions. Here, we report on multi-epoch X-ray observations of Cassiopeia A (Cas A), a 350-year-old SNR, in the 15-50 keV band that probes the most energetic CR electrons. The observed X-ray flux decrease ($15\pm1\%$ over 10 years), contrary to the expected $>$90\% decrease based on previous radio, X-ray, and gamma-ray observations, provides unambiguous evidence for CR electron acceleration operating in Cas A. A temporal model for the radio and X-ray data accounting for electron cooling and continuous injection finds that the freshly injected electron spectrum is significantly harder (exponential cutoff power law index $q=2.15$), and its cutoff energy is much higher ($E_{cut}=36$ TeV) than the relic electron spectrum ($q=2.44\pm0.03$, $E_{cut}=4\pm1$ TeV). Both electron spectra are naturally explained by the recently developed modified nonlinear diffusive shock acceleration (mNLDSA) mechanism. The CR protons producing the observed gamma rays are likely accelerated at the same location by the same mechanism as those for the injected electron. The Cas A observations and spectral modeling represent the first time radio, X-ray, gamma ray and CR spectra have been self-consistently tied to a specific acceleration mechanism -mNLDSA– in a young SNR.

\end{abstract}

\keywords{Supernova remnants(1667) --- Galactic cosmic rays(567) --- X-ray astronomy(1810)}


\section{Introduction} \label{sec:intro}

Supernova remnants (SNRs) have been considered excellent candidates for Galactic cosmic ray (CR) accelerators due to the large energy of supernova explosions ($E_{SN}\gtrsim10^{51}$ erg) and formation of a strong shock (Mach number $\mathcal{M}\gg1$). In particular, at the early stage of their evolution ($<$ 1 kyr), their fast shock velocities ($v_{sh}\sim$ several thousand km s$^{-1}$) and amplified magnetic fields ($B\sim$ a few hundred $\mu$G) make SNRs ideal CR accelerators. The resultant CR spectrum bears crucial information about the shock acceleration mechanism operating in SNRs and about their acceleration environments, such as shock velocity, magnetic field, and ambient matter density. The spectrum of the most energetic CR electrons with teraelectronvolt (TeV, $10^{12}$ eV) energies can be probed via the X-rays they emit by gyrating around magnetic fields (synchrotron radiation). Long-term monitoring with \chandra{} revealed that some young SNRs exhibited localized year-scale increases and decreases of soft X-ray flux in the 4-6 kiloelectronvolt (keV, $10^3$ eV) band by $\sim$ 50\%. This rapid and extreme variability was attributed to fast electron acceleration and synchrotron cooling in a high magnetic field, $B\geq$ 100 $\mu$G \citep{rxj1713,chandra1,tycho}. However, the energetics and spectrum of CR electrons cannot be inferred from such narrow-band observations. A further complication arises due to the contamination of their soft X-ray spectrum by significant bremsstrahlung radiation of thermal electrons ($kT$ $\sim$ a few keV).

A direct measurement of CR electrons comes from hard X-ray observations above $\sim15$ keV where the contamination by thermal electrons is minimal. The spatial distribution of the most energetic CR electrons in Cassiopeia A (Cas A), a young ($\sim350$ years old \citep{age}) nearby (3.4 kpc away \citep{distance}) SNR, was first resolved by the Nuclear Spectroscopic Telescope Array (\nustar{}), a space-based X-ray telescope sensitive in the 3-79 keV band, with a 14$''$ (full width at half maximum) angular resolution. Cas A is a bright X-ray source whose synchrotron emission from ultra-relativistic CR electrons extends up to $\sim50$ keV. The 2.4 Ms of data collected in 2012-2013 showed that X-rays above 15 keV are predominantly emitted from knots \cite{nustar} coincident with the regions that showed fast variability in the soft X-ray observations with \chandra{} \citep{chandra1,Chandra2,Chandra3}. In addition, these regions are located at the reverse shock rather than the forward shock, where the strongest particle acceleration is expected (Figure \ref{fig:diff} (a)). 

\nustar{} observed Cas A again in 2023 for 188 ks. Combined with archival observations (Table \ref{tab:obs}), the multi-epoch \nustar{} data above 15 keV obtained over a 10-yr baseline allowed us a unique opportunity to track the most extreme particle acceleration and cooling process operating in Cas A. 

%%%%%

\section{Multi-epoch X-ray analysis} \label{sec:xray}

\subsection{Observations and data reduction} \label{subsec:data}

Cas A was observed by \nustar{} in August 2012 - December 2013 for a net exposure of 2.2 Ms in total, and in March - April 2023 for a net exposure of 188 ks in total (Table \ref{tab:obs}). We reduced the data using NuSTAR Data Analysis Software (NuSTARDAS) version 2.1.2 and CALDB version 20240325. The NuSTARDAS pipeline produces cleaned event files with good time intervals after screening the South Atlantic Anomaly (SAA) passages and applying data quality cuts. We applied the most conservative criteria for filtering the SAA passages to ensure the most stable and accurate flux measurement.

\begin{table}[h!]
\begin{center}
\caption{List of archival and new observations}\label{tab:obs}%
\begin{tabular}{@{}ccccc@{}}
\toprule
Observation ID & Date & Position angle (deg) & Offset$^{\dagger}$ (arcmin) & Exposure (ks) \\
\hline
40001019002 & 2012-08-18 & 84 & 0.9 & 291 \\
40021001002 & 2012-08-27 & 76 & 2.0 & 170 \\
40021001005 & 2012-10-07 & 33 & 1.5 & 183 \\
40021002002 & 2012-11-23 & 338 & 0.3 & 271 \\
40021002006$^{*}$ & 2013-03-02 & 249 & 3.9 & 136 \\
40021002008$^{*}$ & 2013-03-05 & 249 & 4.0 & 189 \\
40021003003 & 2013-05-28 & 151 & 3.5 & 198 \\
40021011002 & 2013-10-30 & 7 & 1.8 & 236 \\
40021012002 & 2013-11-27 & 335 & 1.1 & 206 \\
40021015003 & 2013-12-23 & 312 & 1.6 & 137 \\
\hline
\multicolumn{4}{c}{Total exposure (archival data)} & 2,017 \\
\hline
40801003002 & 2023-03-24 & 229 & 2.2 & 92 \\ 
40801013002 & 2023-04-04 & 217 & 2.5 & 95 \\ 
\hline
\multicolumn{4}{c}{Total exposure (new data)} & 188 \\
\botrule
\end{tabular}
\end{center}
\tablecomments{The exposure has been corrected for deadtime and SAA passages. For the archival data, only the observations with exposures over 100 ks are listed.}
\tablecomments{$^{*}$Only these two observations (total exposure 325 ks) were used in this work to represent the archival data to minimize the systematic uncertainties.}
\tablecomments{$^{\dagger}$Offsets were calculated as an angular separation between the pointing coordinates and the center of a circular region with a radius of 30'' encompassing the two bright konts on the west (see Figure \ref{fig:diff}).}
\end{table}

\subsection{Background estimation} \label{subsec:background}

Since Cas A is a bright extended ($\sim 6'$ across) source, no region in the \nustar{} field of view ($13'\times13'$) is truly source-free to be used for background estimation. Instead, we modeled the background of each observation using nuskybgd, a code for simulating \nustar{}'s background \citep{nuskybgd}. \nustar{}'s background is comprised of three components: (1) focused cosmic X-ray background (CXB), (2) non-focused (``stray light") CXB, and (3) internal background composed of a continuum and multiple lines. (3) is predominant above 10 keV, while (2) is strongest below 10 keV \citep{nustar}. nuskybgd generates a model for each component and normalizes it utilizing regions outside of the source where no additional emission to the background components is expected. To account for the smearing of Cas A's emission into the regions outside of the remnant due to the finite size of the PSF ($14''$ FWHM), we added a phenomenological source model to the nuskybgd background model. After normalizing, the source model was removed, and the background model was used to simulate the background spectrum in the source region (radius $3'$ circle) with the \texttt{fakeit} command in Xspec for an exposure of 10 Ms. Background images were also generated using the normalized background-only model.

\begin{figure}[t]
\centering
\includegraphics[width=0.8\textwidth]{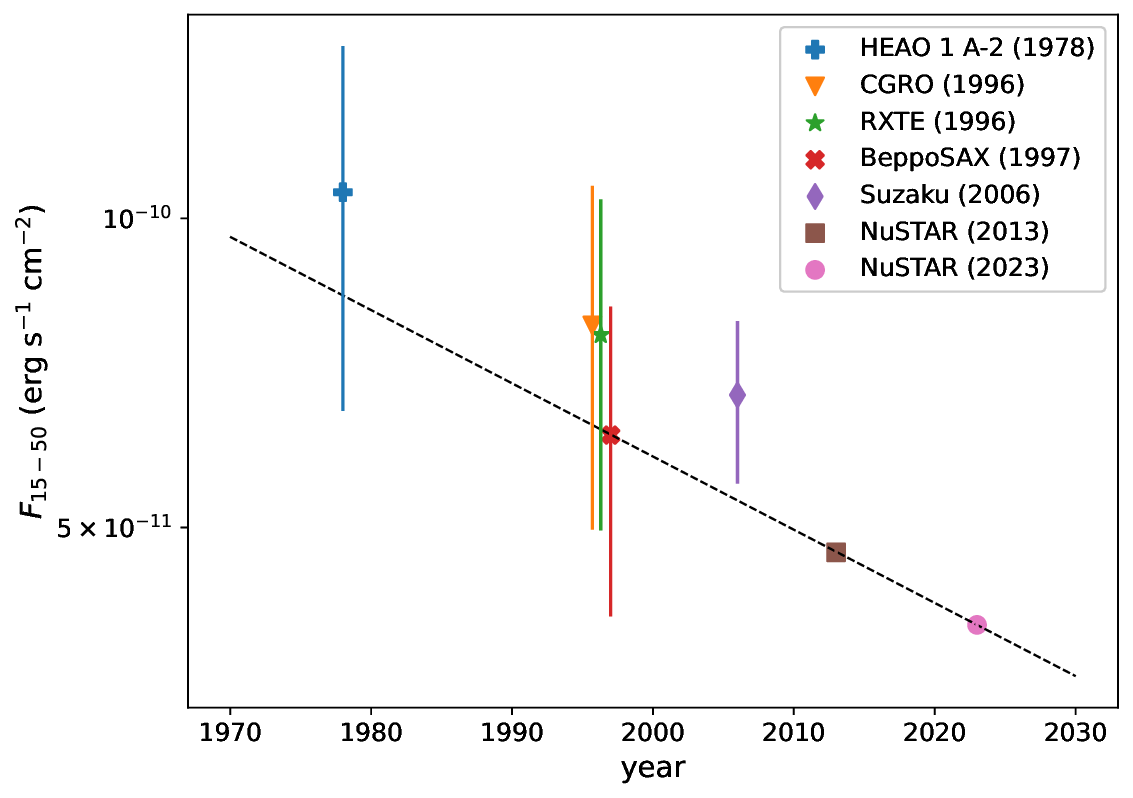}
\caption{\textbf{Historical X-ray flux of Cas A in the 15-50 keV band.} The flux was calculated from the best-fit spectral model for each telescope data. The errors for the spectral model parameters were propagated to calculate the error for the flux. For HEAO 1 A-2, a power law model was fitted to the spectral data reported in \cite{heao}. Previously reported best-fit spectral parameters were used for CGRO \citep{cgro}, RXTE \citep{RXTE}, BeppoSAX \citep{BeppoSAX}, and Suzaku \citep{suzaku}. For \nustar{}, the error bars are smaller than the markers. The dashed black line shows the flux decrease rate ($1.6\pm0.1$)\% yr$^{-1}$ found from a linear regression of all the data points accounting for the error bars.}\label{fig:history}
\end{figure}

\subsection{Spectral analysis} \label{subsec:spectral}

We analyzed the data in the 15-50 keV band. This choice of energy range ensures negligible contribution of thermal X-rays ($kT$ $<4$ keV \citep{temperature1,temperature2,temperature3,temperature4,temperature5}) and source emission above the background level. In addition, the uncertainty of background estimation is minimized in this energy range since the predominant background component (internal background) is well understood, and the smearing of the source counts is minimal.
While combining multiple archival observations may reduce statistical uncertainty in spectral shape and flux measurement, it can introduce even greater systematic uncertainties caused by instrumental (different telescope pointing) and physical (source variability) effects. To minimize these systematic uncertainties, we selected two representative archival observations (observation ID 40021002006 and 40021002008, total exposure 324 ks) to compare with the new observations, given their similar position angles to the new observations and the minimal time gap (3 days) between them. The two selected archival observations are referred to as the archival observations hereafter.

We extracted the source spectra for the entire remnant (redius $6'$ circle). We modeled the spectra with an absorbed power law where the hydrogen column density was fixed to $N_H=2.14\times10^{22}$ cm$^{-2}$ \citep{nh}. The abundance table from \cite{wilm} was used. The spectral parameters were linked among the observations within each epoch. A cross-normalization term was multiplied to each spectra with respect to the FPMA spectra of the earliest observations in each epoch. The average cross-normalizations are $1.019\pm0.005$ for the archival data, and $1.011\pm0.007$ for the new data. The model fits the data well (reduced $\chi^2$ $\sim$ 1). The best-fit power-law photon index for the archival observations is $3.42\pm0.02$, and the 15-50 keV flux is $(4.73\pm0.04)\times10^{-11}$ erg s$^{-1}$ cm$^{-2}$. The new observations after 10 years show that the spectral shape remains unchanged within statistical uncertainties (power-law index $3.37\pm0.02$), and the 15-50 keV flux ($(4.02\pm0.04)\times10^{-11}$ erg s$^{-1}$ cm$^{-2}$) decreased by $(15\pm1)$\% (Figure \ref{fig:history}). If the flux decreased simply by electrons' synchrotron cooling, the spectrum should have become softer due to the inverse relation between synchrotron loss and electron energy. The null-detection of X-ray spectral softening, therefore, indicates that additional physical processes are operating in addition to synchrotron cooling.

For the readers who intend to combine the archival observations of Cas A into a single data point, we present the estimated systematic uncertainties on the spectral measurements for such a case\footnote{While a detailed study of instrumental systematic uncertianties in spectral measurement was performed in \cite{calibration} using the Crab nebula, its result is not directly applicable to our study of Cas A. The study in \citep{calibration} treated the Crab nebula as a point-like source and provided the systematic uncertainty due to a different off-axis angle in the range of 1-$7'$ (1-$4'$ for flux measurement). On the other hand, Cas A is an extended source with a highly inhomogeneous flux distribution; not only an off-axis angle but also a position angle becomes a source of systematic uncertainty in this case. Moreover, depending on the telescope pointing, part of Cas A ($\sim6'$ in diamaeter) may lie beyond the range of an off-axis angle covered by the study in \cite{calibration} for flux measurement ($4'$). Note also that the systematic uncertianties may include the intrinsic flux change of Cas A.}. We analyzed all the archival observations with exposures greater than 100  ks (Table \ref{tab:obs}). We analyzed each observation separately, and linked the parameters between the FPMA and FPMB spectra of each observation using a cross-normalization term with respect to the FPMA. We report the standard deviations of the distributions of power-law index (0.6\%), flux in 15-50 keV (4\%), and cross-normalization (1.2\%) as the systematic uncertainties in case the archival observations are combined into a single data point.

\subsection{Image production} \label{subsec:image}

\begin{sidewaysfigure}
\centering
\includegraphics[width=1.1\textwidth]{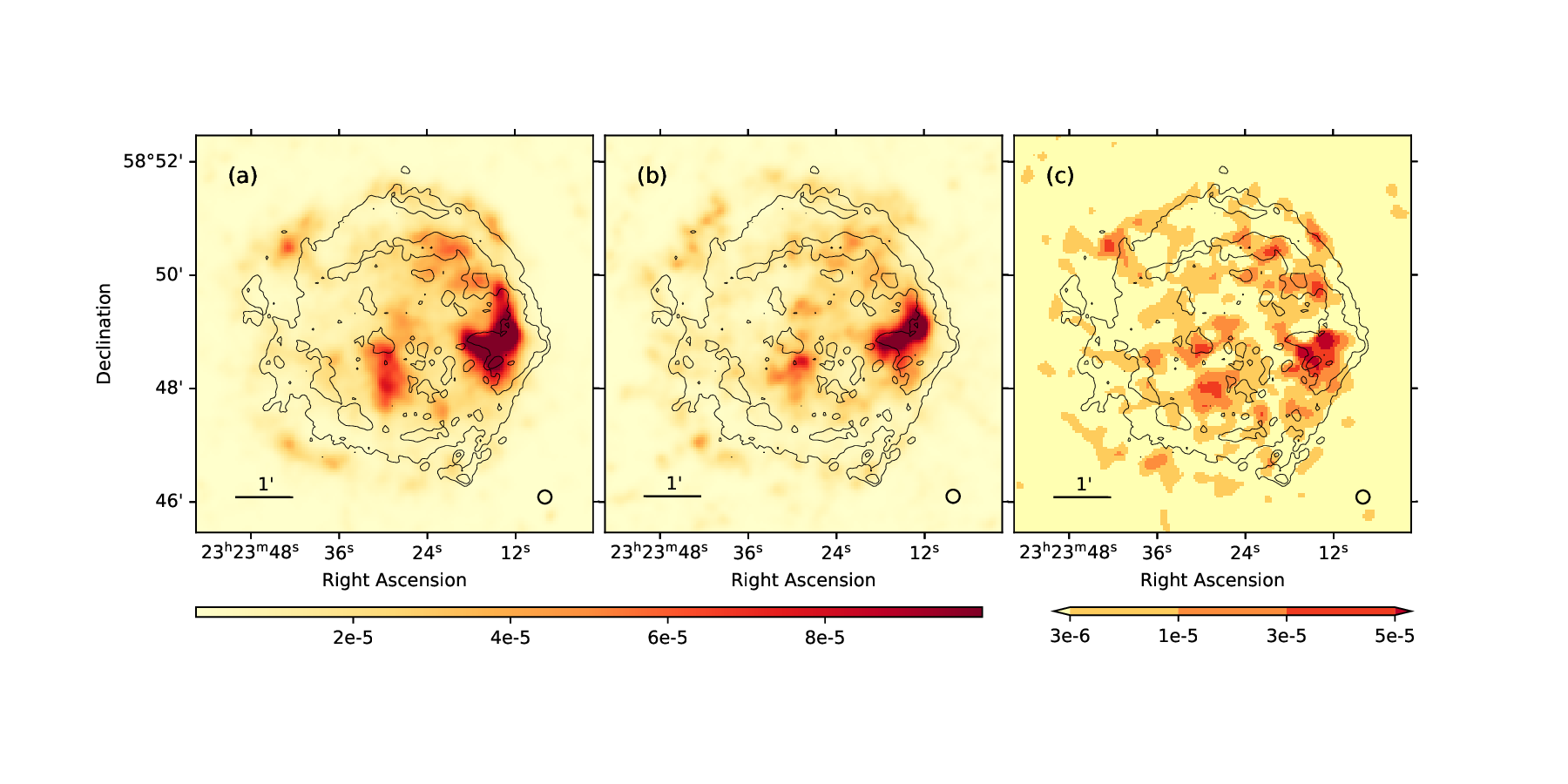}
\caption{\textbf{15-50 keV flux maps of Cas A in 2013 (a) and 2023 (b), and a difference map between 2013 and 2023 (c).} For (a) and (b), background-subtracted counts maps were divided by vignetting-corrected exposure maps, then iteratively deconvolved using the on-axis \nustar{} point spread function (PSF, see \S \ref{subsec:image}). The two images were scaled equally to ensure direct comparison. (c) was generated by subtracting (b) from (a). All three images were smoothed with a Gaussian kernel ($\sigma=3.5''=1.5$ pixel). The L-band radio contour \citep{radio_IR} is overlaid in black. A circle with diameter $14''$ is overlaid at the right bottom corner of each image to indicate the size of the \nustar{} PSF ($14''$ FWHM). Fading over the entire remnant is clearly noticeable. Most prominently, the brightest knots inside the remnant (in the middle and on the west) dimmed significantly.}\label{fig:diff}
\end{sidewaysfigure}

We produced a counts map, background image, and exposure map for each observation and focal plane module in the 15-50 keV range. Vignetting was corrected in the exposure maps for the mean energy (32.5 keV). Flux maps were calculated by subtracting a background image from a count map and dividing it by an exposure map. Individual flux maps were combined within each epoch to generate a flux map for 2013 and 2023. We applied the Lucy-Richardson deconvolution algorithm \citep{Richardson,Lucy} to the flux maps for 50 iterations using the on-axis \nustar{} PSF for the 20-79 keV range (Figure \ref{fig:diff} (a) and (b)).

%%%%%

\section{Temporal synchrotron spectral energy distribution modeling} \label{sec:sed}

For the sub-petaelectronvolt (PeV, $10^{15}$ eV) electrons emitting synchrotron X-rays with photon energy $E_{\gamma}$ in a highly amplified magnetic field $B$ (0.1-1 mG), \cite{bfield1,bfield2,bfield3,bfield4,bfield7,bfield5,bfield6}, the synchrotron cooling time \citep{Tcool}
\begin{equation}
    t_{1/2}=12\left(\frac{E_{\gamma}}{\textrm{10 keV}}\right)^{-1/2}\left(\frac{B}{\textrm{100 } \mu\textrm{G}}\right)^{-3/2}\textrm{yr}
\end{equation}
is much shorter than the length of our baseline. Without ongoing electron acceleration, the hard X-ray flux of Cas A would have decreased by more than 90\%. Instead, we observe a $(15\pm1)$\% decrease in the 15-50 keV flux from the whole remnant over the past 10 years (Figure \ref{fig:history}, uncertainties are 1-$\sigma$ hereafter). The largest flux decrease is observed at the bright knots on the western rim of the reverse shock (Figure \ref{fig:diff} (c)). Moreover, the synchrotron cooling mechanism naturally softens the radiation spectrum (producing a larger photon index $\Gamma$ when $dN_{\gamma}/dE_{\gamma}\propto E_{\gamma}^{-\Gamma}$), whereas our spectral analysis does not find a statistically significant spectral softening in the 15-50 keV band ($\Gamma=3.42\pm0.02$ in 2013 and $\Gamma=3.37\pm0.02$ in 2023). Extreme electron acceleration must be operative in Cas A injecting freshly accelerated electrons. The X-ray spectrum of the injected electrons, when combined with the X-ray spectrum of the rapidly cooling preexisting electrons, leads to an essentially constant photon index with time.

To constrain the spectrum of ongoing electron acceleration in Cas A, we modeled a multiwavelength spectral energy distribution (SED) of Cas A. We first constructed the multiwavelength SED of the whole remnant using the \nustar{} spectrum from this work and the radio spectrum from \cite{radio} for each epoch. The radio spectrum was calculated for each epoch in the L (1395 MHz) and X (9000 MHz) band using the temporal spectral model in \cite{radio} (eq. 14, Table 5). This radio spectral model provides the best fit to the 20-yr (1995-2014) Green Bank Observatory (GBO) 40-foot L-band data and the 3-yr (2014-2017) GBO 20-m L-band and X-band data. A Gaussian quadrature sum of the 1-$\sigma$ uncertainty of each model parameter was used as a 1-$\sigma$ uncertainty of the radio spectrum (eq. 15, Table 5). The radio flux variability calculated from this model is 8\% and 6\% decrease in the L and X bands, respectively.

\subsection{SED model description} \label{subsec:sed_description}

The SED was modeled with a synchrotron radiation spectrum in a Gaussian turbulent magnetic field \citep{gaussB}. A recent X-ray polarization measurement by the Imaging X-ray Polarimetry Explorer (IXPE) \citep{IXPE} suggests magnetic turbulence in Cas A on a scale smaller than $\sim0.4$ pc (IXPE angular resolution $24''$ at the source distance 3.4 kpc). A Gaussian distribution of the magnetic field strength in an SNR is theoretically motivated (e.g., \cite{gaussB_SNR}). The magnetic field distribution is assumed unchanged between 2013 and 2023. Nonthermal electrons present in 2013 (``preexisting electrons") are modeled with an exponential cutoff power law distribution:
\begin{equation}
    \frac{dN}{dE}=N_0\left(\frac{E}{\textrm{1 TeV}}\right)^{-q}\textrm{exp}\left(-\frac{E}{E_{cut}}\right)^{\beta}.
\end{equation}
The minimum and maximum energy bound for the electron distribution were set to $E_{min}=100$ MeV and $E_{max}=3$ PeV, respectively. The electrons lose energy by adiabatic and synchrotron cooling every time step.

An injected electron spectrum follows an exponential cutoff power law distribution with a distinct set of parameters from preexisting electrons. $N_0$ is determined by normalizing the distribution to (constant injection rate) $\times$ (time step) between the same $E_{min}$ and $E_{max}$ as preexisting electrons. $\beta=2$ is held fixed to reflect the synchrotron-loss-limited electron acceleration at the Bohm limit \citep{ZA07}. These electrons are injected and lose energy every time step by adiabatic and synchrotron cooling. 

The same adiabatic and synchrotron loss formulae are used for both preexisting and injected electrons. Adiabatic energy loss is
\begin{equation}
\dot{E}_{ad}=\frac{\langle\dot{R}_{sh}\rangle}{\langle R_{sh}\rangle}E,
\end{equation}
where $E$ is an electron energy, $\langle\dot{R}_{sh}\rangle/\langle R_{sh}\rangle$ is an average expansion rate of all electrons. Pitch-angle-averaged synchrotron energy loss is \citep{gaussB}
\begin{equation}
    \langle\dot{E}_{syn}\rangle=\frac{4}{3}\sigma_T\left(1-\frac{1}{\gamma^2}\right)\gamma^2c\frac{B^2_0}{8\pi},
\end{equation}
where $\sigma_T$ is the Thomson cross-section, $\gamma$ is an electron Lorentz factor, and $B_0$ is a standard deviation of a Gaussian distribution of magnetic field strength. Synchrotron loss of radio-emitting electrons is negligible ($\ll1\%$) over the 10-year period of our consideration for any reasonable magnetic field strength $<$ a few mG. The adiabatic loss rate $\sim0.3\%$ yr$^{-1}$ is necessary to reproduce the observed radio spectral variability. This rate is comparable to the average expansion rate of Cas A's forward shock ($0.218\pm0.029$)\% yr$^{-1}$ measured with multi-epoch \chandra{} observation \citep{expansion}. On the other hand, for X-ray-emitting electrons, synchrotron loss ($\gg50\%$) overpowers adiabatic loss in any reasonable magnetic field strength $>$ a few tens of $\mu$G. 

We first find a preexisting electron spectrum that reproduces the \nustar{} and radio spectrum in 2013. Then an injected electron spectrum is added every time step on top of the best-fit preexisting electron spectrum, and both electron populations are cooled every time step to reproduce the \nustar{} and radio spectrum in 2023. The size of the time step is min(0.1 yr, $E_{max}/\dot{E}_{syn}$). 

\subsection{SED modeling results} \label{subsec:sed_results}

\begin{figure}[t!]
\centering
\includegraphics[width=0.49\textwidth]{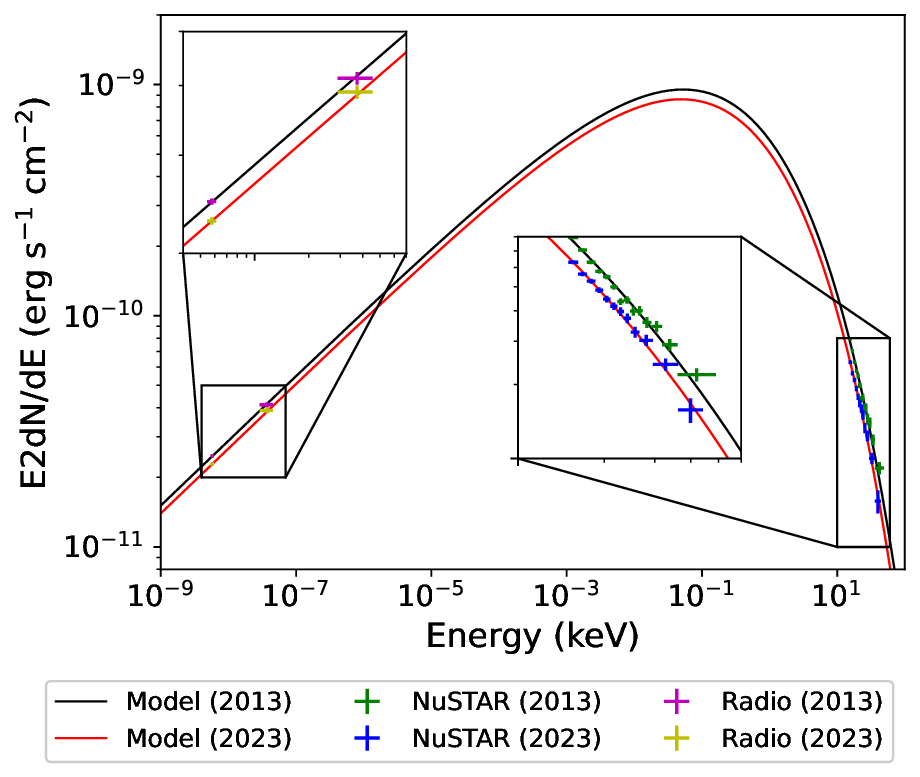}
\includegraphics[width=0.5\textwidth]{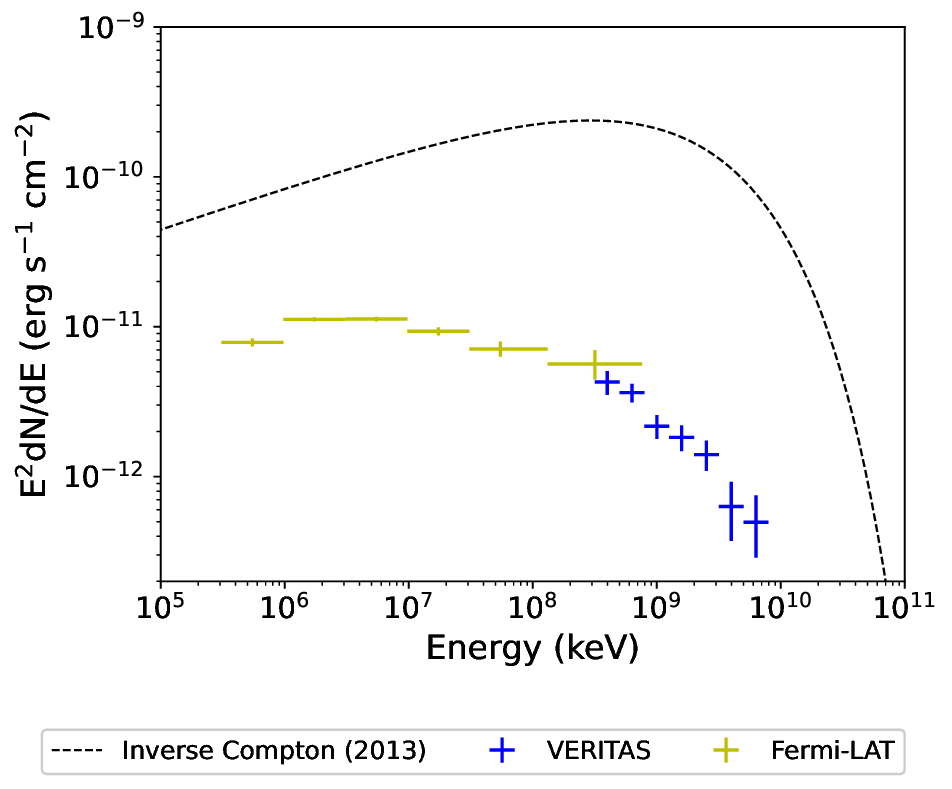}
\includegraphics[width=0.51\textwidth]{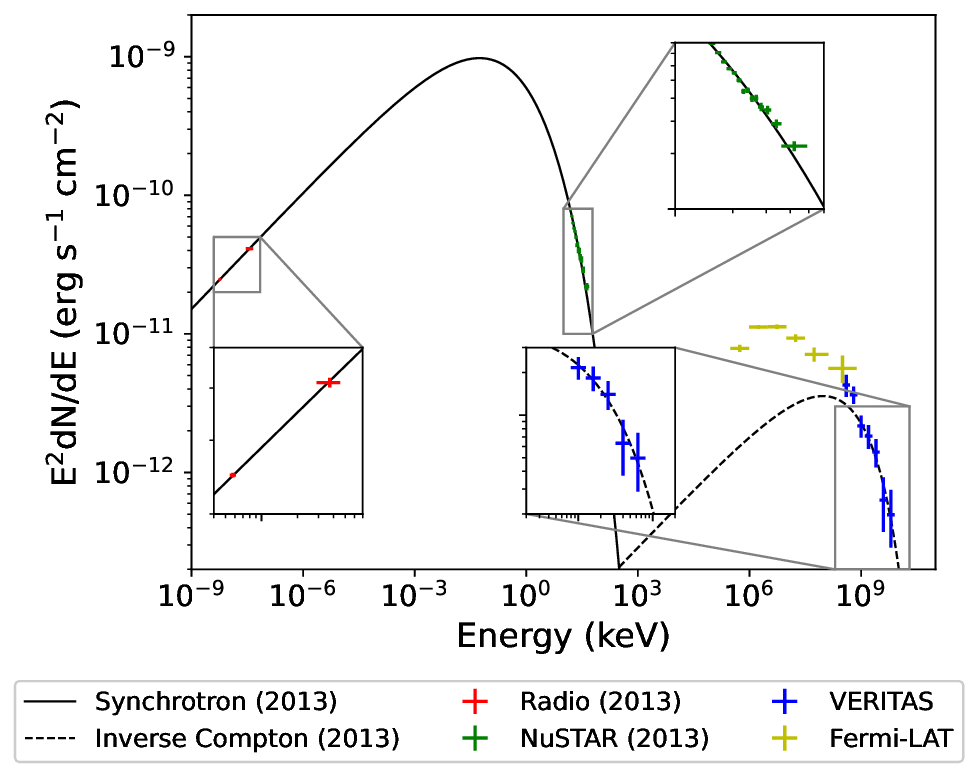}
\includegraphics[width=0.48\textwidth]{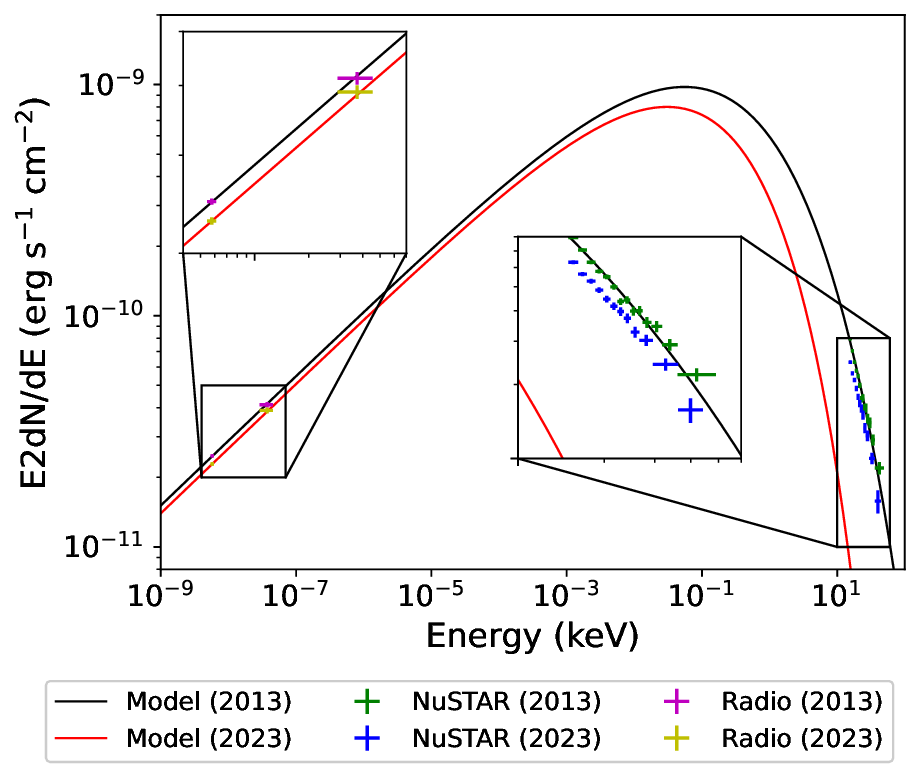}
\caption{\textbf{Top: Synchrotron (left) and inverse Compton (right) spectrum from the best-fit electron distribution for $B_0=6$ $\mu$G.} With this magnetic field, the predicted gamma-ray flux significantly exceeds the observed gamma-ray flux. \textbf{Bottom: Synchrotron and inverse Compton spectrum in 2013 (left), and the synchrotron spectrum in 2013 and 2023 (right) from the best-fit electron distribution for the lower-limit magnetic field $B_0=123\pm3$ $\mu$G.} The lower-limit magnetic field predicts over 50\% decrease in the 15-50
keV flux, much faster than the observed decrease ($15\pm1$\%). The \fermi{} and VERITAS spectra were taken from 4FGL-DR4 \citep{4FGL,4FGL-DR4} and \cite{veritas2}, respectively.}\label{fig:cool}
\end{figure}

Setting the injection rate to zero, magnetic field $B_0\sim6$ $\mu$G is required for the observed 15\% flux decrease in the 15-50 keV band (Figure \ref{fig:cool} top left). This magnetic field is at least an order of magnitude smaller than the previous estimations \citep{bfield1,bfield2,bfield3,bfield4,bfield7,bfield5,bfield6}. Inverse Compton scattering \citep{ICS} of the cosmic microwave background and infrared (temperature $\sim100$ K, energy density 2 eV cm$^{-3}$ \citep{IR}) photons off the electron distribution found with this magnetic field produces significantly more gamma rays than those observed by VERITAS \citep{veritas2} and \fermi{} \citep{4FGL,4FGL-DR4} (Figure \ref{fig:cool} top right), or MAGIC \citep{magic}.

Assuming that the observed TeV gamma rays are produced entirely by inverse Compton Scattering of electrons, the lower limit on the magnetic field is found to be $B_0=123\pm8$ $\mu$G (Figure \ref{fig:cool} bottom left). This is comparable to the lower limit placed by \cite{veritas2} ($\sim150$ $\mu$G) using a similar approach. The best-fit electron distribution found with this magnetic field, however, reproduces the radio and X-ray spectrum only in 2013. Due to a much faster synchrotron cooling than when $B_0=6$ $\mu$G, the predicted X-ray flux in 2023 is $\sim10$ times lower, and the spectrum is much softer than the observed X-ray flux. This discrepancy between the prediction and observation leaves no possibility other than electron injection into Cas A compensating extremely rapid energy loss of preexisting electrons. 

We constrain the characteristics of the preexisting and injected electron spectra using the lower-limit magnetic field found in this work ($123\pm8$ $\mu$G) and the upper-limit magnetic field established in the previous works ($B\sim1$ mG, \cite{bfield1,bfield2,bfield3,bfield4,bfield7,bfield5,bfield6}). The best-fit model parameters for preexisting and injected electrons for magnetic fields $B_0=(123, 400, 1000)$ $\mu$G are listed in Table \ref{tab:model}. \\

\begin{table}[h!]
\begin{center}
\caption{Best-fit parameters for the synchrotron cooling and injection model}\label{tab:model}
\begin{tabular}{c|c|ccc}
\toprule
Electron population & Preexisting electrons$^{\dagger}$ & \multicolumn{3}{c}{Injected electrons} \\
\hline
$B_0$ ($\mu$G) & $123\pm8$ & $123\pm8$ & 400$^{\ast}$ & 1000$^{\ast}$ \\
\hline
$q$ & $2.44\pm0.03$ & $2.2$ & $2.0$ & $1.9$ \\
$E_{cut}$ (TeV) & $4\pm1$ & 36 & 19 & 12 \\
$\beta$ & $0.74\pm0.05$ & 2$^{\ast}$ & 2$^{\ast}$ & 2$^{\ast}$ \\
Total electron energy (fraction)$^{\ddagger}$ & $(1.09\pm0.02)\times10^{49}$ erg & 0.4\% & 0.9\% & 2.1\% \\
Injection rate ($10^{37}$ erg s$^{-1}$) & $-$ & 14 & 4 & 2 \\
\botrule
\end{tabular}
\end{center}
\tablecomments{$^{\ast}$Parameters were held fixed.} 
\tablecomments{$^{\dagger}$The parameters for preexisting electrons are given only for the lower-limit magnetic field $B_0$ $=123\pm8$ $\mu$G. For higher magnetic fields, $q$ and $\beta$ do not change, while $E_{cut}$ and total electron energy decrease.} 
\tablecomments{$^{\ddagger}$Total electron energy is reported for preexisting electrons. The fraction of the total energy of the preexisting electrons that were injected between 2012-2013 is reported for injected electrons.}
\end{table}

%%%%%

\section{Discussion}

\subsection{Hard injection spectrum} \label{subsec:injection}

\begin{figure}[h!]
\centering
\includegraphics[width=0.6\textwidth]{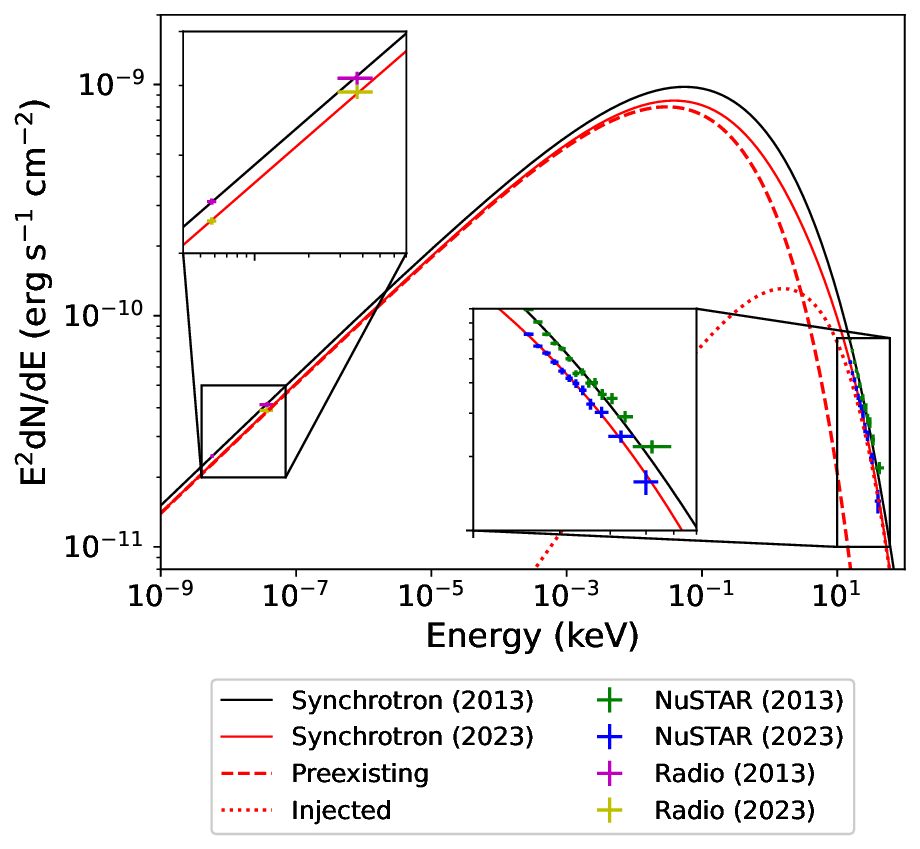}
\caption{\textbf{Synchrotron spectrum in 2013 and 2023 calculated with the temporal synchrotron cooling and injection model.} The black (red) solid line is the synchrotron spectrum calculated using the best-fit parameters in Table \ref{tab:model} for 2013 (2023). The red dashed line is the synchrotron spectrum of the preexisting electrons in 2023, and the red dotted line shows the contribution of the injected electrons to the synchrotron spectrum in 2023. The \nustar{} spectrum from this work and the radio spectrum from \cite{radio} are overlaid. The error bars of the data points are for $1\sigma$ uncertainties in this figure and all the figures hereafter.}\label{fig:cool_inject}
\end{figure}

For the lower-limit magnetic field of $B=123$ $\mu$G, 0.4\% of the total energy of the preexisting electrons needs to be injected over 10 years. The corresponding electron spectrum ($dN/dE\propto E^{-q}\textrm{exp}[(-E/E_{cut})^{\beta}]$)
is significantly harder for the injected electrons ($q=2.15$) than that of the preexisting electrons ($q=2.44\pm0.03$). The cutoff energy of the spectrum is much higher for the injected electrons ($E_{cut}=36$ TeV) than the preexisting electrons ($E_{cut}=4\pm1$ TeV) (Figure \ref{fig:cool_inject}). 

For the upper-limit magnetic field of $B\sim1$ mG, 2.1\% of the preexisting electron energy needs to be injected with an even harder electron spectrum ($q=1.86$), and cutoff energy is still higher than the preexisting electrons ($E_{cut}=12$ TeV). The higher magnetic field produces faster synchrotron cooling, and that requires more electron injection, while the electron spectral index $q$ and the cutoff energy $E_{cut}$ are self-consistently regulated by the radio and \nustar{} data in 2023, respectively.

Further evidence that such hard-spectrum injection has been operating in Cas A comes from radio and infrared observations. Spectral hardening in the radio band has been observed since 1949 \citep{radio2,radio3,radio,radio4}. Spectral hardening continues in the infrared band as the infrared synchrotron flux was measured to exceed the power-law extrapolation of the radio flux \citep{infrared,infrared2}. The radio-infrared spectral hardening is most significant at the bright X-ray knots where the hard-spectrum injection is expected to operate \citep{radio_IR}. 

The preexisting electron spectrum reflects the entire history of electron acceleration in Cas A up to 2013. Since low-energy electrons have a cooling time scale much longer than the age of Cas A, the spectral index of the preexisting electrons is governed by the synchrotron radio spectrum of low-energy electrons. On the other hand, the injected electron spectrum reflects only the electron acceleration operating in Cas A at the current time. This is mainly determined by the fast-cooling high-energy electrons and their synchrotron hard X-ray variability. 

Equivalently, the hard X-ray morphology locates only the current CR acceleration site (injected electrons), whereas the radio morphology also reveals the location of CRs accelerated in the past (preexisting electrons). The hard X-ray morphology of Cas A shows a high concentration of emission on the knots. On the other hand, the radio morphology is more isotropic with enhancement at the same knots and along the reverse shock (e.g. \cite{radio_IR}). The brightening in both wavelengths locates the knots as the current most active particle acceleration sites (e.g. \cite{reverse_shock}), while the globally isotropic radio emission traces the electrons that were accelerated at an earlier time, most likely at the forward shock, and diffused out.

%Therefore, distinct sets of environmental parameters, such as magnetic field, shock speed, and plasma density, need to be adopted when connecting the preexisting and injected electron spectrum to their acceleration mechanism.

The soft-spectrum electrons that were accelerated earlier dominate the radio-emitting electron population, whereas the hard-spectrum electrons that were accelerated recently dominate the X-ray-emitting electron population. The combination of these two distinct electron populations creates a slower cutoff in the overall electron spectrum (best-fit cutoff index $\beta=0.74\pm0.05$), compared with analytic solutions for synchrotron-loss-limited diffusive shock acceleration in the case of Bohm diffusion ($\beta=2$ \citep{ZA07}). This slow cutoff in the electron spectrum propagates to the synchrotron spectrum creating a harder X-ray spectrum than the aforementioned synchrotron-loss-limited case. There were efforts to explain this spectral behavior of Cas A by hard-spectrum ($q=2.1$) electrons accelerated at the fast-moving jet-like structure \citep{jet} or the jitter radiation (magnetobremsstrahlung emission of electrons in magnetic turbulence with a scale much smaller than the electron gyroradii) \citep{jitter1,jitter2,jitter3,jitter_CasA}. The former overpredicts the upper limits placed by LHAASO in 10 TeV - 1 PeV \citep{lhaaso}, and the latter requires a magnetic turbulence scale $<100$ km, much smaller than an observable scale (e.g., IXPE angular resolution $24''$ is equivalent to 0.4 pc $=3\times10^{13}$ km at the distance 3.4 kpc). Our work provides the most natural explanation for the observed spatial-dependent spectral behavior of Cas A.

%%%%%

\subsection{Modified nonlinear diffusive shock acceleration} \label{subsec:mnldsa}

One can examine if the derived CR spectra are consistent with recent theoretical expectations. Nonlinear diffusive shock acceleration (NLDSA) theories address CR-driven modification to the standard DSA by introducing a region upstream of a shock with an enhanced density due to CR pressure (``precursor", e.g. \citep{NLDSA1,NLDSA2,NLDSA3,NLDSA4}) that leads to an increased shock compression ratio ($R>4$). Recent hybrid (kinetic ion-fluid electron) simulations \citep{mnldsa1,mnldsa2} discovered that the equivalent of a precursor is formed downstream of a shock (``postcursor") whose contribution to the increased compression ratio dominates that of a precursor. On the other hand, the compression ratio experienced by CRs is decreased since magnetic fluctuations and CRs drift away from the shock in a postcursor. This postcursor effect on the ``effective" compression ratio is parameterized by a factor $\alpha$, which depends on shock velocity, upstream density, and downstream magnetic field. In mNLDSA, the correction factor $\alpha$ for shock compression ratio can be deduced from observables as \citep{mnldsa2}
\begin{equation}\label{eq:alpha}
    \alpha \simeq 5 \frac{B}{\textrm{1 mG}} \frac{\textrm{1000 km s}^{-1}}{v_{sh}} \left(\frac{R}{5} \frac{\textrm{cm}^{-3}}{n}\right)^{1/2},
\end{equation}
where $B$ is a post-shock magnetic field, $v_{sh}$ is a shock speed, $R$ is a shock compression ratio, and $n$ is a pre-shock plasma number density. Measuring $R$ is nontrivial. Adopting a nominal $R=5$ (e.g. $R\sim4-7$, \citep{SN1006}) and observed $v_{sh}\sim6000$ km s$^{-1}$ \citep{inward-chandra,expansion}, $n\sim1$ cm$^{-3}$ \citep{density}, and $B=0.7$ mG (0.1 mG $\lesssim B \lesssim$ 1 mG), eq. \ref{eq:alpha} gives $\alpha\sim0.6$.

Within this framework (modified NLDSA or mNLDSA), the CR spectral index is given by $q=3R/(R-1-\alpha)-2$. Typical values of $R\sim5$ and $\alpha\sim0.6$ \citep{mnldsa2} predict $q\sim2.4$ as we found for the preexisting electron population. At the reverse shock, our inferred spectrum of $q\sim2.2$ is consistent with the theoretical prediction if the postcursor is not present (in which case $\alpha\rightarrow0$ and $R\rightarrow4$). This is plausible since, at the reverse shock, the postcursor may not form due to the presence of contact discontinuity in the downstream region.

\subsection{Proton acceleration at the hard X-ray knots?} \label{subsec:proton}

The injected electron spectrum ($q\sim2.2$) agrees with the proton spectrum found from gamma-ray observations ($q\sim2.2$, \cite{veritas2}), indicating that the same acceleration mechanism is operating for both electrons and protons at the same acceleration site. The hard X-ray knots at the reverse shock are observed to move inward at a much higher speed than the forward shock ($v_{sh}\sim8000$ km s$^{-1}$ in the ejecta frame, \cite{inward-chandra,expansion}) while the rest of the reverse shock is still moving outward. As it has been proposed, this requires the presence of an overdense region on the western rim of the reverse shock, such as molecular clouds \citep{CasA-MC,inward-shock} or an asymmetric circumstellar shell \citep{reflected}. CR protons accelerated at the reverse shock can inelastically scatter with the overdensity and produce copious gamma rays by pion production and decay. 

%%%%%

\section{Summary and conclusion} \label{sec:summary}

Our multi-epoch hard X-ray observations enabled the isolation of a pure synchrotron radiation component of Cas A. Such radiation is associated with energetic, nonthermal CR electrons. By temporal modeling of this emission, we were able to establish the existence of two distinct populations of CR electrons. One population is more energetic and associated with a powerful, active accelerator, and the second population is less energetic and associated with an accelerator that was more active in the past. The X-ray morphology of Cas A allowed us to identify the sites of both accelerators.  These observations and associated modeling provide the first self-consistent analysis of a young SNR that connects the CR spectrum and acceleration location to the broadband multiwavelength spectrum from radio to gamma-ray energies with interpretation through the most recent work on the mNLDSA theory. Our observational and theoretical approach can be applied to other young SNRs to elucidate the acceleration mechanism and environment of CRs below the ``knee" of their spectrum at $\sim3$ PeV.

%% Also note that the akcnowlodgment environment does not support long amounts of text. If you have a lot of people and institutions to acknowledge, do not use this command. Instead, create a new \section{Acknowledgments}.
\begin{acknowledgments}
We thank T. Sato, D. Caprioli, R. Diesing, and S. Reynolds for the discussions. Support for this work was partially provided by NASA through \nustar{} Cycle 8 Guest Observer (GO) Program grant NNH21ZDA001N. J.W. was partially supported by NSF grant WOU MMA PHY-2110497.
\end{acknowledgments}


\vspace{5mm}
\facilities{\nustar{}}


\software{NuSTARDAS, Xspec \citep{xspec}, NumPy \citep{numpy}, SciPy \citep{SciPy}, Astropy \citep{astropy:2013,astropy:2018,astropy:2022}, Gammapy \citep{gammapy:2023,gammapy1.1}, Matplotlib \citep{matplotlib}, Naima \citep{naima}}




\clearpage


\bibliography{main}{}
\bibliographystyle{aasjournal}



\end{document}